# Cultural noise and the night-day asymmetry of the seismic activity recorded at the Bunker-East (BKE) Vesuvian Station

Nicola Scafetta[1*], Adriano Mazzarella[1]

[1] Department of Earth Sciences, Environment and Georesources, University of Naples Federico II, Largo S. Marcellino, 10 - 80138 Naples, Italy

* nicola.scafetta@unina.it

## Abstract

Mazzarella and Scafetta (2016) showed that the seismic activity recorded at the Bunker-East (BKE) Vesuvian station from 1999 to 2014 suggests a higher nocturnal seismic activity. However, this station is located at about 50 m from the main road to the volcano's crater and since 2009 its seismograms also record a significant diurnal cultural noise due mostly to tourist tours to Mt. Vesuvius. Herein, we investigate whether the different seismic frequency between day and night times could be an artifact of the peculiar cultural noise that affects this station mostly from 9:00 am to 5:00 pm from spring to fall. This time-distributed cultural noise should evidently reduce the possibility to detect low magnitude earthquakes during those hours but not high magnitude events. Using hourly distributions referring to different magnitude thresholds from M = 0.2 to M = 2.0, the Gutenberg–Richter magnitude-frequency diagram applied to the day and night-time sub-catalogs and Montecarlo statistical modeling, we demonstrate that the day-night asymmetry persists despite an evident disruption induced by cultural noise during day-hours. In particular, for the period 1999-2017, and for earthquakes with M ≥ 2 we found a Gutenberg–Richter exponent b = 1.66 ± 0.07 for the night-time events and b = 2.06 ± 0.07 for day-time events. Moreover, we repeat the analysis also for an older BKE catalog covering the period from 1992 to 2000 when cultural noise was not present. The analysis confirms a higher seismic nocturnal activity that is also characterized by a smaller Gutenberg–Richter exponent b for M ≥ 2 earthquakes relative to the day-time activity. Thus, the found night-day seismic asymmetric behavior is likely due to a real physical feature affecting Mt. Vesuvius.

# 1. Introduction

Mazzarella and Scafetta (2016) have recently analyzed the two main seismic catalogs of Mt. Vesuvius, which were collected at the Osservatorio Vesuviano Ovest (OVO) and at Bunker-East (BKE) Vesuvian (Osservatorio Vesuviano, http://www.ov.ingv.it). The downloadable OVO catalog covered the period from Feb/23/1972 to Nov/11/2014, while the downloadable BKE catalog covered the period from Jan/01/1999 to Sep/07/2014 (http://www.ov.ingv.it/ov/en/banche-dati/186-catalogo-sismico-del-vesuvio.html). According to the legends of the two published catalogs: OVO collected 11,804 seismic events with magnitude between −0.4 and 3.6 and has completeness threshold of magnitude $M \geq 1.9$ (cf.: Luongo et al., 1996a; Duma and Vilardo, 1998; Duma and Ruzhin, 2003); BKE collected 11,042 seismic events with magnitude between −2.5 and 3.6 but has a far better completeness threshold: $M \geq 0.2$ (Orazi et al., 2013). Thus, the location and the higher seismic sensitivity of the BKE station make it more useful to monitor Mt. Vesuvius. Indeed, BKE is just about 1.1 km away from the center of the volcano's crater while OVO is about 2.6 km far away.

Mazzarella and Scafetta (2016) found that both OVO and BKE catalogs evidence the existence of a higher nocturnal seismic activity occurring at the Mt. Vesuvius. The result confirmed and extended that of Duma and Vilardo (1998) who analyzed the seismic activity recorded at OVO station from 1972 to 1996 and showed that this seismic activity was characterized by a diurnal oscillation with maxima during night-hours. We concluded that the existence of a diurnal cycle could suggest influences due to the meteorological diurnal cycle typically found in temperature, humidity and wind stress records, or due to the diurnal cycle of the geomagnetic field or perhaps due to a gravitational effect related to the S1 tidal harmonic. For example, Neuberg (2000) and Scafetta and Mazzarella (2015) and Mazzarella and Scafetta (2016) suggested an external modulation of seismic activity driven by meteorological factors and climate oscillations.

However, in principle there might be an objection and a simple explanation for the found diurnal seismic cycle at Mt. Vesuvius: namely, a significant day-time cultural noise due mostly to heavy vehicles bringing tourists to visit the volcano's crater. This phenomenon has become relevant since 2009.

Cultural noise is detected by the seismographs and could hide low magnitude earthquake events. Thus, despite the BKE catalog up to 2014 was claimed to be complete for events with a magnitude of $M \geq 0.2$, numerous minor events could have been missed during day-times. As a consequence, day-time cultural noise could generate a catalog of seismic events characterized by an apparent higher nocturnal activity simply because night-hours are free of cultural noise.

Herein, we will show that, indeed, the BKE station is severely affected by day-time cultural noise because extremely close to the main road used by several buses to bring tourists to visit the volcano's crater. However, using (1) hourly distributions referring to different magnitude thresholds, (2) the Gutenberg–Richter magnitude-frequency diagram and (3) Montecarlo statistical modeling, we demonstrate that the day-night asymmetry in the BKE seismic catalog persists despite the evident disruption induced by day-time cultural noise. We refer to Mazzarella and Scafetta (2016) for our previous analysis and extended comments that are not repeated here.

Moreover, to further confirm our result we also found and herein analyze an older BKE catalog covering the period from Feb/1992 to Dec/2000 (Sarao et al., 2001). This catalog is important because, at that time, cultural noise around the BKE station was not relevant and the seismograms were very clean also during day-times. We show that frequency analysis and Gutenberg–Richter diagrams applied to this older catalog confirm the night-day seismic asymmetric behavior affecting Mt. Vesuvius found in the 1999-2014 catalog, which was the one analyzed in Mazzarella and Scafetta (2016) .

However, while this paper was under review, the Osservatorio Vesuviano has updated the BKE record up to October 2017. We add a short section in order to analyze also the updated record that now covers the period from Jan/01/1999 to Oct/1/2017.

## 2. Description of the cultural noise at the Bunker-East (BKE) Vesuvian Station

The Bunker-East Vesuvian Station (lat. 40°49'.07 N; log. 14° 26'.33 E; elev. 863 m slm) is located on the East dorsal of Mt. Vesuvius: see Figure 1. This site records the seismicity due to the volcano's activity since February 1992 (Buonocunto et al., 2001). The general technical characteristics of the BKE seismic station are found in Sarao et al. (2001).

As Figure 1 shows, BKE is located at about 50 m from Matrone road. Since 2009, touristic tours to the Mt. Vesuvius crater are organized daily, usually from 9:00 am to 5:00 pm except in winter from January to March, when weather conditions are good (http://busviadelvesuvio.com). The tourist transport occurs through specific middle-size ecological green buses that can carry up to 25 tourists at a time that drive along the Matrone road, one of the oldest tracks of Mt. Vesuvius. The vehicles stop at a small parking at the end of the road which is located at an elevation of about 1050 m slm, passing by the location of the BKE station. Then, from that parking the tourists reach the crater by walking along the Gran Cono pathway that also encircles the crater: see Figure 1.

The buses along the Matrone road approach the location of the BKE station making the ground to vibrate. This cultural noise is easily detected by the seismograph. Figure 2 shows its severity by depicting a BKE seismogram referring to a 24-hour interval from 9:00 pm on May/03/2017 to 9:00 pm on May/04/2017 in local times. The depicted 24-hour seismogram is typical for the days when touristic tours to Mt. Vesuvius occur.

Figure 2 shows that the signal is completely noise-free during the night-hours. However, during the day-hours the seismic signal is severely disrupted by vehicles induced tremors. In fact, the seismogram clearly detects a vibration that gradually increases as a bus approaches BKE on the Matrone road and then gradually decreases. Figure 1 also shows that, above the BKE station, Matrone road rises up the volcano turning left and right by forming a "S" shape pathway. Thus, for every one-way trip to or from the parking at 1050 m slm a vehicle passes right above the BKE station three times at different altitudes, as indicated by the intersections of the road with the yellow segment depicted in Figure 1. This dynamics generates a typical tremor made of three main beats as it is observed in Figure 2.

In particular, Figure 2 shows a tremor occurred from 6:42 to 6:46 am likely due to the passage of a cross-country vehicle of the park officers. This tremor appears smaller than the following ones because this vehicle was evidently lighter than the touristic buses. Probably, park officers check in the early morning the condition of the road because landslides are frequent on Mt. Vesuvius. At about 8:53 am the first touristic bus arrives: note the three-beats signal with the first beat to be the strongest one (= the bus is coming to the volcano). At about 4:55 pm the last bus leaves the volcano: note the three-beats signal with the third beat to be the strongest one (= the bus is leaving).

For a proper evaluation of the disruption of the seismogram induced by cultural noise, its signatures must be compared against real earthquake signatures so that the former can be distinguished from the latter. Figure 3 shows an example of a BKE seismogram that depicts a number of real earthquake events with their relative magnitude, which occurred on Aug/08/2011 from 0:00 to 4:00 am GTM (1:00 am to 5:00 am local time).

A real earthquake signature clearly differs from that of a vehicle passing nearby a seismograph. In the case of a vehicle, the signal gradually increases, reaches a maximum and then gradually decreases. On the contrary, for a real earthquake the intensity of the fluctuations increases very fast, nearly instantaneously, indicating a sudden fracture in the ground and then it decreases slowly. Thus, an expert technician can easily distinguish the two types of signatures.

By comparing Figures 2 and 3, it is possible to qualitatively conclude that earthquakes with a magnitude between 1 and 1.5 could be easily covered by bus-induced noise for at least 2-3 minutes at each bus-passage. Because there are about 30-35 bus passages, on 12 day-time hours the blind interval for middle size events could be at most 100 minutes on a total of 720 minutes. This means that about 15% of day-time earthquake events could be missing during the tourist seasons. This percentage could have a seasonal modulation given that more tourists visit the volcano in summer than in spring or fall while this kind of tours are not organized in winter.

The percent of missing events would increase for very small earthquakes with M < 1.0. To quantify this situation, from Figure 2 it is possible to measure that each bus-related seismic signal lasts about 7 minutes; this means that for about 30-35 bus passages up to 245 on 720 day-time minutes would be blinded by the cultural noise and the catalog would be expected to miss about 35% of all very small magnitude earthquakes during the tourists seasons.

On the contrary, Figures 2 and 3 suggest that earthquakes with a magnitude $M \geq 2.0$ could be unlikely covered by the bus-induced cultural noise for more than about 30 seconds at each bus passage. Thus, the blind interval could be less than 20 on 720 minutes. This means that less than 3% of the large magnitude earthquake events could be missed during day-hours. Moreover, earthquakes with a magnitude $M \geq 1.9$ are easily detected at OVO. Thus, it would be very unlikely for an expert technician to miss earthquakes with a large magnitude despite the cultural noise.

In conclusion, while it is possible that during the tourist seasons the BKE catalog might miss several day-time small magnitude earthquake events, it should be considered essentially complete for the large magnitude ones, at least for $M \geq 2.0$.

In any case, although the above considerations are quite qualitative and the above statistics of potentialy missing event is just indicative, one of the two following logical conclusions would be expected:

1) if the earthquake events were randomly and uniformly distributed in the 24 hours, the night-time and day-time frequencies should become closer to each other for stronger earthquakes;

2) if, on the contrary, a higher frequency of nocturnal seismic activity recorded at Mt. Vesuvius persists or even increases for stronger earthquakes, the result of Mazzarella and Scafetta (2016) should be considered robust despite the presence of a day-time cultural noise effecting the BKE station during specific periods.

The following analysis will check the issue.

## 3. The 1999-2014 BKE database

3.1. Analysis of the hourly histograms at varying magnitude thresholds.

The BKE catalog can be directly downloaded from the OV web-site (http://www.ov.ingv.it). Figure 4 depicts all events from Jan/01/1999 to Oct/01/2017 that include those from Jan/01/1999 to Sep/07/2014 studied by Mazzarella and Scafetta (2016), whose magnitude is reported in the catalog, by distinguishing them in blue and red dots whether they occur during night-times or day-times, respectively. These periods are defined below.

A simple visual inspection of Figure 4 already suggests that for M ≥ 2.0 many more earthquakes occurred during night-time than day-time.

We explicitly quantify the relative frequencies by calculating the histograms of the hourly probability of earthquakes using different magnitude threshold levels from M ≥ 0.2, that is the BKE claimed magnitude completeness level, to M ≥ 2.0. Figure 5 shows that during the night-time the earthquake frequency is significantly greater than during the day-time at all considered magnitude thresholds. As explained in section 2, if the detected higher nocturnal seismic activity were due to a cultural noise artifact, the difference between night-time and day-time frequency should decrease as the magnitude threshold increases. However, the contrary is observed with the night-day asymmetry becoming more prominent when stronger magnitude thresholds are selected.

Figure 5 also permits to identify the best night-time and day-time periods when their statistical difference is maximum. We found that such maximum occurs when we select as the day-time hours the interval from 7:00 am to 6:59 pm, and as the night-time hours the period from 7:00 pm to 6:59 am.

Table 1 reports the probability of day-time and night-time earthquake events during several time intervals as the magnitude threshold level increases from M ≥ 0.2 to M ≥ 2.0.

Under the null hypothesis that earthquake events are evenly distributed during night and day times, the third column of Table 1 indicates the percent of hypothetical missing events during day-time. This factor is calculated as 100(1-df/nf), where df is the day-time frequency and nf is the night-time frequency. For example, under such a null hypothesis during day-time about a third (~33%) of events with M ≥ 0.2 would be missing in the catalog because of the day-time cultural noise.

First, we note that such 33% of hypothetical missing events is unlikely because from Section 2 such a value would be about the maximum possible percent of daily missing events for M ≥ 0.2 when cultural noise is present. However, the real maximum percent of missing events should be significantly less, about 10%, because the analyzed record covers the period 1999-2014 while the touristic-related cultural noise is present since 2009, that is for 6 on 16 years, and for just 9 months a year.

Moreover, according the above null hypothesis, as the magnitude threshold increases, the missing event probability should decrease. However, Table 1 (third column) shows that it increases to 41% for M ≥ 1.7 events and to 46% for M ≥ 2.0. From the discussion developed in Section 2, it is statistically very unlikely that the observed cultural noise would be able to obscure larger earthquakes more often than the smaller ones. Thus, the result indicates that during night-times a larger number of stronger earthquakes did occur at Mt. Vesuvius during the period covered by the catalog despite cultural noise might have obscured many small magnitude events during the day-times.

This result can be double checked by noting (see Figure 2) that cultural noise is present roughly for only 8 hours from 9:00 am (when the first touristic bus arrives) to 4:59 pm (when the last touristic bus leaves) while the day-time hours from 7:00 am to 8:59 am and from 5:00 pm to 6:59 pm are mostly free of noise. Thus, the catalog could be missing a certain percent of events occurring only from 9:00 am to 4:59 pm alone. Under the null hypothesis that the earthquake events are evenly distributed during the 24 hours, the probability to record earthquakes from 7:00 am to 8:59 am and from 5:00 pm to 6:59 pm should be compatible with that observed during the night-hours from 7:00 pm to 6:59 am of the following day.

The sixth and seventh column of Table 1 compares the hourly average probabilities of earthquake events for the 8-hour period from 9:00 am to 4:59 pm, and the 4-hour period from 7:00 am to 8:59 am and from 5:00 pm to 6:59 pm. We observe that the probability of earthquake during the 4 noise-free day-time hours is always significantly lower than during the night-hours: in the former case the mean hourly probability of events is ≤ 3.58% while in the latter case is ≥ 4.92%.

In this regards, the earthquake event probability during the 4 noise-free day-time hours is closer to that referring to the 8 noisy day-time hours. We find that the occurrence of M ≥ 0.2 to M ≥ 1.3 earthquake events is slightly greater during the 4 noise-free day-time hours than during the 8 noisy day-time hours: 3.48-3.58% vs. 3.21-3.36%, respectively. However, the two probabilities are equal for M ≥ 1.5 earthquake events (3.35%) and the probability for the 4-hour interval gets even lower than that for the 8-hour period for M ≥ 1.7 to M ≥ 2.0 earthquake events: 2.26-2.99% vs. 3.13-3.31%, respectively. Thus, also this result further confirms that during night-times a stronger number of larger earthquakes did occur at Mt. Vesuvius from 1999 to 2014.

3.2. Interpretation of the Gutenberg–Richter diagram

One of the most useful relations adopted in seismology is the Gutenberg–Richter law (Gutenberg and Richter, 1954). This empirical law establishes that the magnitude and total number of earthquakes in any given region and time period of at least that magnitude are related according to the following equation:

$$N = 10^{a-bM} = N_{TOT}\, 10^{-bM}, \qquad [1]$$

where N is the number of events having a magnitude ≥ M and a and b are two constants. The a-value is herein of a less scientific interest because it just indicates the total seismicity rate of the region: $N_{TOT} = 10^a$. On the contrary, $10^{-bM}$ is the occurrence probability for events with a magnitude ≥ M. In reality, the trend expressed by Eq. 1 is the tail of a cumulative lognormal distribution (see Paparo and Gregori, 2003). A similar consideration refers to the Omori law referring to the time interval distribution of earthquakes (cf. Scafetta and West, 2004).

Because catalogs usually miss earthquakes as their magnitude decreases and for very strong earthquakes their number becomes increasingly insufficient for statistical analysis, real Gutenberg–Richter distributions are characterized by a roll-off for very small magnitudes and a saturation for very large magnitudes. Thus, for practical purposes, the coefficient b of Eq. 1 is estimated using a normalized cumulative probability function with $N/N_{TOT}$ vs. M in a log-linear diagram that is fit with a straight line of the type f(x) = a-bx in the interval characterized by a linear behavior.

Among earthquake catalogs the coefficient b is found to be on average close to 1. However, some spatial and temporal variation for the b-value is also observed in the approximate range of 0.5 to 2.0, which can become as large as 2.5 during earthquake swarms. The b-value characterizes the seismic physical properties of a region (Luongo et al., 1996a, 1996b).

About the interpretation of the magnitude of the coefficient b, we note that the physical system looks for its new equilibrium state after having suffered by an irreversible process that led to a rupture of its solid structure, i.e. after having suffered by a catastrophe. An area of high b-values could be indicative of a relative low stress regime while an area of low b-values can be interpreted as evidence of a relatively higher stress regime associated with an area of dominantly extensional stress (Farrell et al., 2009). It may also be said that low b-values may refer to areas characterized by deeper earthquakes, where the stress is usually higher, while large b-values may refer to areas characterized by more shallow and, therefore, brittle crustal components. Thus, because higher values of b indicate a smaller probability that very large earthquakes could occur frequently, such a coefficient can be used to characterize the seismic hazard of a region.

In particular, volcanic regions such as Mt. Vesuvius should be characterized by high b-values for strong events (cf: Sarao et al., 2001) because usually these earthquake events may be quite shallow, e.g. they may occur within the volcano cone edifice itself. For example, using the Mt. Vesuvius earthquake record covering the period from October 2016 to September 2017 downloadable from the OV web-site, we found that on 703 total events about 50% occurred at a depth lower than 220 m, 80% occurred a depth lower than 500 m, and 91% occurred a depth lower than 1000 m, Since Mt. Vesuvius is 1280 m high slm, the great majority of its seismic events occur within its volcano cone edifice. Moreover, the several fractures present in the shallow volcanic grounds make unlikely the accumulation of the large stress required for large magnitude earthquakes. Note that the number of earthquakes is related to the size of a fractured zone according to a power law showing a fractal dimension equal to b (Utsu, 1969).

Figure 6 depicts the Gutenberg–Richter diagrams regarding the cumulative probability functions for all BKE events (red), for the night-time alone (blue) and for the day-time alone (green). Two different linear ranges are observed for events with magnitude 0.0 ≤ M ≤ 2.0 and for 2.2 ≤ M ≤ 3.0. The ranges with M ≤ 0 and M ≥ 3.0 are not considered for estimating the coefficient b because of roll-off and statistical scarcity.

As Figure 6 shows, for the magnitude interval $0.0 \leq M \leq 2.0$, the Gutenberg–Richter exponent is $b = 0.78 \pm 0.02$ and is common for all three records. On the contrary, for the magnitude interval $2.2 \leq M \leq 3.0$ we found $b = 1.69 \pm 0.05$ for all events, $b = 1.58 \pm 0.07$ for the night-time events and $b = 1.94 \pm 0.06$ for day-time events.

To interpret the importance of the result, first we notice that under the hypothesis of physical homogeneity, the normalized Gutenberg–Richter cumulative distribution of earthquakes is independent on the total number of earthquakes. Thus, if a catalog is randomly cut in half, the normalized Gutenberg–Richter cumulative distribution would still be equal for the two halves and equal to the original entire distribution except for a statistical fluctuation that may generate a small difference.

Thus, under the null hypothesis that earthquake events are evenly distributed during night and day times, the night-time and day-time subsets of events should still be characterized by a similar Gutenberg–Richter cumulative distribution even in the case that a random number of day-time earthquake events were missing in the catalog due to the presence of a day-time cultural noise.

However, for the night-time earthquake set, we find a Gutenberg–Richter b-exponent smaller than that measured for the day-time earthquake set. This suggests that the physical condition of the night-time seismicity of Mt. Vesuvius is different than that observed during day-time, at least for events with magnitude $M \geq 2.0$. This day-night asymmetry is not limited to only the number of observed events, as showed in the hourly histograms depicted in Figure 5, but also to their very physical characteristic as measured by their respective Gutenberg–Richter b-exponents.

Let us discuss another statistical test to demonstrate the significance of the above result. Let us again assume the null hypothesis that earthquake events are evenly distributed during night and day times but that the day-time catalog is partially incomplete because several small and medium magnitude earthquakes have been obscured by cultural noise. Thus, we can hypothetical assume that the real day-time earthquake sequence is statistically equivalent to that observed at BKE during night-time, which is made of 6991 events. Then, we assume that the real earthquake sequence is disrupted by cultural noise in such a way that, for example, only $M \geq 2.0$ events are not disrupted as argued in section 2, while a number of the $M \leq 2.0$ events are randomly covered by the cultural noise. Finally, we impose the condition that the produced fictitious day-time catalog has a number of events equal to that observed at BKE during day-time, that is made of 4473 events for all reported magnitudes. We built this statistics using a Montecarlo algorithm generating 10000 alternative synthetic day-time catalogs with the above properties.

Figure 7 compare the Gutenberg–Richter cumulative distributions for the events observed during night-times (blue) and during day-times (green) versus the average Gutenberg–Richter cumulative distribution (red) for the synthetic day-time catalogs produced as described above with relative 1-σ error bars generated by the Montecarlo algorithm.

We observe that the Gutenberg–Richter cumulative distribution referring to the synthetic catalogs (red) is above that referring to the generating sequence (blue) for $M \geq 1.3$. The two distributions are nearly identical only for $M \leq 1.3$. By construction, for $M \geq 2.0$ the synthetic record has a distribution higher than the generating one by a constant factor equal to $6991/4473 = 1.56$. On the contrary, the Gutenberg–Richter cumulative distribution referring to the real day-time sequence (green) is clearly below both the blue and the red curves also considering the error bars of the model.

This evidence is magnified in the insert of Figure 7 that depicts the ratio between the synthetic day-time Gutenberg–Richter cumulative distribution and the night time Gutenberg–Richter cumulative distribution with relative error bars (red), and the ratio between the real day-time Gutenberg–Richter cumulative distribution and the night time Gutenberg–Richter cumulative distribution (green). Here it is explicitly shown that the real day-time distribution follows the prediction of the synthetic one only up to $M \leq 1.3$ and then the divergence becomes very evident with the ratio green curve falling well below 1 while the synthetic one rises to a 1.56 ratio value above to the night-time distribution.

The above Montecarlo experiment can be repeated using a large magnitude threshold per events not affected by the noise, but the qualitative result would remain the same. Namely the red curve representing the Gutenberg–Richter cumulative distribution for the Montecarlo synthetic records representing the hypothetical day-time catalogs would be on average always above the blue curve representing the Gutenberg–Richter cumulative distribution for the night-time events while the green curve for the real day-time catalog is clearly below it.

## 4. The 1992-2000 BKE database

Sarao et al. (2001) published an early seismic BKE catalog made of 9003 events covering the period from Feb/1992 to Dec/2000, that is since the BKE's installation. It is unknown to us why these data are not included in the BKE catalog that can be downloaded from the OV web-site that, as state above, starts in Jan/01/1999.

A direct comparison between the two catalogs for the overlapping period from Jan/01/1999 to Dec/31/2001 gives 1991 and 1175 events for the years 1999 and 2000 for the old catalog while for the new one there are 1992 and 1328 events for the same years, respectively. Thus, 254 additional events are recorded in the new catalog, nearly all for the year 2000. Also the reported magnitude sometime differs in the two catalogs. Thus, the new catalog was likely based on revised methodologies to interpret the seismograms, but this revision probably started just in 1999, which perhaps is the reason why OV publishes on its web-site only the catalog starting since 1999. Thus, it is better not to merge the two catalogs but to consider them two independent measure of the seismic activity on Mt. Vesuvius since (1) they cover different periods that overlap for just two years, and (2) the two catalogs were apparently collected in a different way.

Sarao et al. (2001)'s catalog is relevant for our case because from 1992 to 2000 cultural noise due to touristic visits to the volcano was nearly absent, at least from Matrone road. As explained above, this kind of cultural noise started since 2009. Thus, we repeat the analysis to test whether the found night-day seismic asymmetric behavior persists also during this early period.

Sarao et al. (2001)'s catalog is depicted in Figure 8. We separate the events occurred during the night-hours (7:00 pm to 6:59 am, local time) from those occurred during the day-hours (7:00 am to 6:59 pm, local time). A simple inspection and comparison between Figures 8a and 8b immediately evidence that during the night-time 8 events with $M \geq 3.1$ occurred while no earthquake with an equivalent large magnitude occurred during day-time.

These eight events are reported in Table 3. The probability that such an occurrence could happen by chance in the case the earthquake events were randomly distributed between day and night times is just P = $100*0.5^8$ = 0.4%.

Table 2 analyzes the 1992-2000 BKE catalog. The table reports the cumulative percent of events in function of the magnitude occurred during the night-hours (7:00 pm to 6:59 am) and during the day-hours (7:00 am to 6:59 pm) and their relative variation calculated as 100(1-df/nf) that indicates the percent of how less events are recorded during the day time relative to the night-time.

The reported values confirm that not only there are more events during the night-time relative to the day-time but that the discrepancy increases when stronger earthquakes are selected. For example, for the entire catalog and for events with M ≥ 0.2, 52% of events occurred during the night-hours and 48% occurred during the day-hours, that is the day-time frequency was 6% lower than the night-time frequency according to the equation 100(1-df/nf), as reported in Table 2. However, for M ≥ 2.0, 57% of events occurred during the night-hours and 43% occurred during the day-hours, that is a 25% difference.

A similar behavior was also observed for the 1999-2014 catalog, although in the latter case the percent variation values between night-time and day-time events were greater likely because of the cultural noise affecting that period. As also explained in the previous section, the increase of the night-day discrepancy contradicts the hypothesis that the reduced number of day-time events could be an artifacts of some form of day-time cultural noise.

The Gutenberg–Richter cumulative distribution depicted in Figure 9 more clearly evidences that stronger earthquakes occur more frequently during night-time than during day-time. As for Figure 7, we find again that for the strongest earthquakes (M ≥ 1.8) the Gutenberg–Richter exponent b is smaller for the night-time events (b = 1.18 ± 0.03) than for the day-time events (b = 1.35 ± 0.04). For earthquake with M < 1.8, the cumulative distributions are identical with a Gutenberg–Richter exponent b = 1.26 ± 0.05.

5. The 1999-2017 updated BKE database

While this paper was under review, the Osservatorio Vesuviano web-site has updated the BKE record from September 2014 to up to October 2017, and the update is now weekly. The new record here analyzed covers the period up to October/01/2017 and contains 2852 additional events (with 144 events without magnitude) for a total of 14656 events since January/01/1999 (with 484 events without magnitude).

The last three years are particularly important because of the heavy day-time cultural noise that has characterized the record. In fact, of the 2852 new events, 1078 occurred during day-time from 7:00 am to 6:59 pm, and 1775 occurred during night-time from 7:00 pm to 6:59 am. For events with magnitude M ≥ 0.2, 33% occurred during day-time and 67% during night time. However, for events with magnitude M ≥ 1.9, still 30% occurred during day-time and 70% during night time. For events with magnitude M ≥ 2.3, still 33% (just 3 events) occurred during day-time and 67% (6 events) during night time. No events with magnitude M ≥ 2.6 has been observed during the period. Thus, again the evidence is that during night-time there are more events that during day-time. In fact, even if the day-time cultural noise could hide about 50% of the total events with M ≥ 0.2 occurred during those hours, it is extremely unlike that the same noise could still hide about 50% of larger magnitude earthquakes with M ≥ 1.9 and M ≥ 2.3 that could be easily detected also at OVO.

The updated 1999-2017 BKE record contains 5848 daytime events (229 events without magnitude) and 8808 nighttime events (255 events without magnitude). Figure 10 depicts the Gutenberg–Richter diagrams regarding the cumulative probability functions for all BKE events (red), for the night-time alone (blue) and for the day-time alone (green) for all records. We found the same behavior already observed in Figure 6 and 9 where, for events with magnitude M ≥ 2.0, the exponent b is significantly larger for the daytime events (b = 2.06 ± 0.07) than for the nighttime events (b = 1.78 ± 0.05). For the magnitude interval 0.0 ≤ M ≤ 2.0, we found a Gutenberg–Richter exponent b = 0.81 ± 0.02 common for all three records.

## 6. Discussion and Conclusion

Mazzarella and Scafetta (2016) found that the seismic activity recorded at the BKE Vesuvian station is characterized by a higher nocturnal activity. However, this station is located at about 50 m from the main road to the volcano's crater and its seismic recording sensitivity is significantly lowered by diurnal cultural noise that is mostly due to buses passing by the BKE station bringing tourists to visit Mt. Vesuvius usually from 9:00 am to 5:00 pm. This problem did not exist when the BKE station was first installed in 1992. This cultural noise problem has become relevant mostly since 2009 when tourist-tours started to be organized using buses driving up to the Mt. Vesuvius along Matrone road passing by the BKE station.

We have used hourly distributions in function of different magnitude thresholds from M = 0.2 to M = 2.0, the Gutenberg–Richter magnitude-frequency diagram applied to the day and night-time sub-catalogs and Montecarlo statistical modeling. We concluded that the observed day-night seismic asymmetry can not be a mere artifact of the significant cultural noise effecting the BKE station during day-time since 2009 from spring to fall.

In particular, we stress the results of the Gutenberg–Richter diagram. For the magnitude interval 0.0 ≤ M ≤ 2.0, we found a Gutenberg–Richter exponent of about b = 0.80 common for all three records. This indicates that for small magnitude earthquakes Mt. Vesuvius behaves mostly as a rigid body. On the contrary, using the record from 1999 to 2014, in the magnitude interval 2.2 ≤ M ≤ 3.0, we found b = 1.69 ± 0.05 for all events, b = 1.58 ± 0.07 for the night-time events and b = 1.94 ± 0.06 for day-time events. Using the updated record from 1999 to 2017, in the magnitude interval 2.2 ≤ M ≤ 3.0, we found slightly larger values: b = 1.78 ± 0.05 for all events, b = 1.66 ± 0.07 for the night-time events and b = 2.05 ± 0.06 for day-time events. This indicates that for large magnitude earthquakes Mt. Vesuvius fractures more easily, although during night-time it can accumulate a higher stress and stronger events might occur more likely.

Our result appears robust because the Gutenberg–Richter diagram is not altered significantly when a certain percent of earthquake events is randomly missing in a catalog. The result has been also confirmed by Montecarlo statistical modeling. Moreover, the fact that the seismic activity on Mt. Vesuvius is higher during night-time is also confirmed by an older BKE catalog covering the period 1992-2000 when cultural noise at BKE due to tourist tours to the volcano did not exist. The difference in the records found for the 1992-2000 and the 1999-2017 BKE catalogs could be explained by the fact that the latter was affected by cultural noise and perhaps was compiled with a revised methodology.

The fact that the Gutenberg–Richter exponent b is higher for day-time events than for night-time events reveals an important seismic physical characteristic of Mt. Vesuvius. In fact, it has been suggested that the observed areal and temporal variations of the Gutenberg–Richter b-values found among the seismic catalogs could be due to a different stress applied to the material (Scholz, 1968), to the depth (Mori and Abercombie, 1997), to the focal mechanism (Schorlemmer et al., 2005), to the strength heterogeneity of the material (Mogi, 1962). In lab experiments it has also been observed that the *b*-value decreases prior to the samples' failure (Lockner and Byerlee, 1991) so that the b-value could be a precursor to major macro-failure (Smith, 1981).

In the case of Mt. Vesuvius the day-night variation of the Gutenberg–Richter b-value should indicate a daily variation of the state of stress of the vulcano's cone edifice. The result suggests influences due to the meteorological diurnal cycle in temperature, humidity and wind stress, or due to the diurnal cycle of the geomagnetic field or perhaps due to a gravitational effect related to the S1 tidal harmonic, or a combination of them as already discussed in Mazzarella and Scafetta (2016).

Mazzarella and Scafetta (2016) also argued for an annual and seasonal variation. In fact, Figure 10 in Mazzarella and Scafetta (2016) shows that a higher nocturnal seismic activity at the Mt. Vesuvius is always observed at an annual and a seasonal scale for M ≥ 0.2 events also from 1999 to 2009. Thus, the result appears robust. However, since 2009 the night-day difference is likely stressed by cultural noise.

Hence, mechanical fracturing occurs on the outermost layers mainly during cooling hours of the day, much more than during the warming hours.

A night/day asymmetry of the seismic activity was observed and reported by Gregori and Paparo (2004) (and subsequent papers by the same authors) on the Gran Sasso by means of acoustic emission (AE) records. The hypothesized cause was the diurnal cycle of warming and cooling rocks. When rocks are heated, the outer layers are first heated and experience thermal expansion over the innermost cooler rocks that are more contracted. When rocks cool, the outer layers cool first, and later on also the inner layers. The outer layer contract due to thermal contraction, while the inner layers are still warmer and more expanded.

Although the real cause of the found night-day seismic asymmetric behavior is still unknown, herein we have definitely demonstrated that it is a real physical feature effecting Mt. Vesuvius and not an artifact due to day-time cultural noise.

**Acknowledgment:** We thank Dr. Luigi Maisto, president of Vesuvian guides for useful discussions.

**Original data available from:**

1999-2017 BKE Catalog, accessed on Oct/03/2017:

http://www.ov.ingv.it/ov/en/component/content/article/75-generale/186-catalogo-sismico-del-vesuvio.html

1992-2000 BKE Catalog, accessed on Oct/03/2017:

http://www.iaea.org/inis/collection/NCLCollectionStore/_Public/33/047/33047973.pdf

| Mag. | Night Time nf | Day Time df | Night-Day variation | Night Average 12 hours | Day average 12 hours | Day average 8 hours | Day average 4 hours |
|---|---|---|---|---|---|---|---|
| M $\geq$ 2.0 | 64% | 36% | 44% | 5.33% | 3.00% | 3.31% | 2.26% |
| M $\geq$ 1.9 | 65% | 35% | 46% | 5.42% | 2.92% | 3.20% | 2.31% |
| M $\geq$ 1.7 | 63% | 37% | 41% | 5.25% | 3.08% | 3.13% | 2.99% |
| M $\geq$ 1.5 | 60% | 40% | 33% | 5.00% | 3.33% | 3.35% | 3.35% |
| M $\geq$ 1.3 | 59% | 41% | 30% | 4.92% | 3.42% | 3.36% | 3.55% |
| M $\geq$ 1.0 | 60% | 40% | 33% | 5.00% | 3.33% | 3.24% | 3.58% |
| M $\geq$ 0.7 | 60% | 40% | 33% | 5.00% | 3.33% | 3.26% | 3.48% |
| M $\geq$ 0.2 | 60% | 40% | 34% | 5.00% | 3.33% | 3.21% | 3.51% |

**Table 1:** The 1999-2014 BKE catalog. First and second column: probability of earthquake occurrence during night (from 7:00 pm to 6:59 am) and during day (from 7:00 am to 6:59 pm) using different magnitude threshold levels (cf. Figure 5). Third column: variation between day and night frequency calculated as 100(1-df/nf). Fourth to seventh column: hourly mean probability for the night and day times, for the 8-hours day-time interval from 9:00 am to 4:59 pm, and for the 4-hour day-time interval from 7:00 am to 8:59 am and from 5:00 pm to 6:59 pm.

| Mag. | night-time nf | day-time df | variation |
|---|---|---|---|
| M ≥ 2.0 | 57% | 43% | 25% |
| M ≥ 1.9 | 57% | 43% | 23% |
| M ≥ 1.7 | 56% | 44% | 21% |
| M ≥ 1.5 | 55% | 45% | 18% |
| M ≥ 1.3 | 53% | 47% | 13% |
| M ≥ 1.0 | 53% | 47% | 11% |
| M ≥ 0.7 | 52% | 48% | 8% |
| M ≥ 0.2 | 52% | 48% | 6% |

**Table 2:** The 1992-2000 BKE catalog. Percent of events using different magnitude thresholds occurred during the night-hours (7:00 pm to 6:59 am), during the day-hours (7:00 am to 6:59 pm) and their relative variation calculated as 100(1-df/nf).

| date | Time | Magnitude | Duration (s) |
|---|---|---|---|
| 1992/08/05 | 4:29 | 3.1 | 100 |
| 1993/01/11 | 22:20 | 3.1 | 100 |
| 1993/01/14 | 3:53 | 3.5 | 135 |
| 1995/08/02 | 2:07 | 3.4 | 100 |
| 1995/09/24 | 2:26 | 3.1 | 120 |
| 1996/04/25 | 20:55 | 3.4 | 120 |
| 1999/10/09 | 5:41 | 3.5 | 130 |
| 1999/10/11 | 2:35 | 3.1 | 100 |

Table 3: The eight largest earthquakes occurred between 1991 and 2000. They all occurred during the night-time.

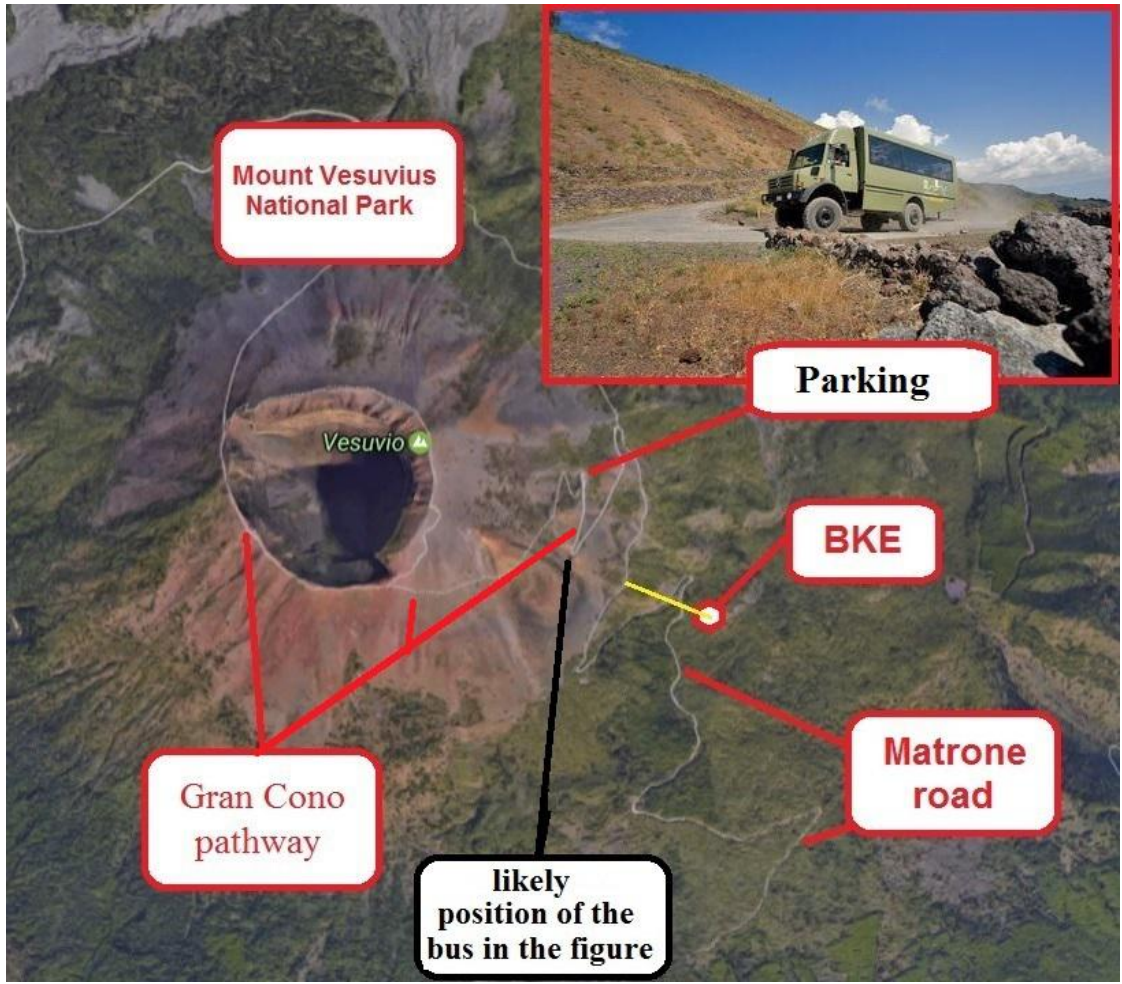

**Figure 1:** Location of the BKE station (lat. 40°49'.07 N; log. 14° 26'.33 E; elev. 863 m slm) nearby the Mt. Vesuvius. In the insert it is shown a touristic bus approaching the parking at 1050 m slm through Matrone road and the Gran Cono pathway.

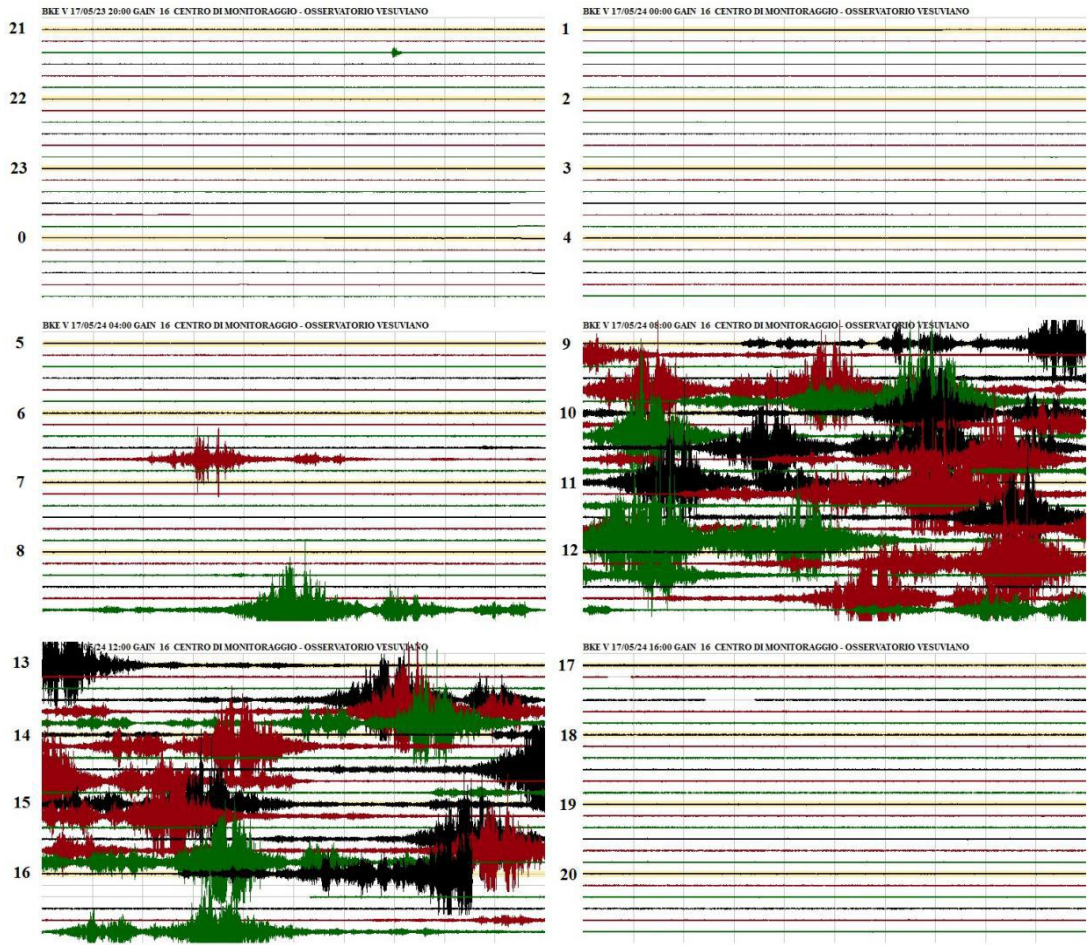

**Figure 2:** An example of a 24-hour BKE seismogram from 9:00 pm (= 21:00) of May/23/2017 to 8:59 pm (= 20:59) of May/24/2017. Each diagram shows a 4-hours record. The local time is indicated at the left side of each diagram. Each line covers 10 minutes.

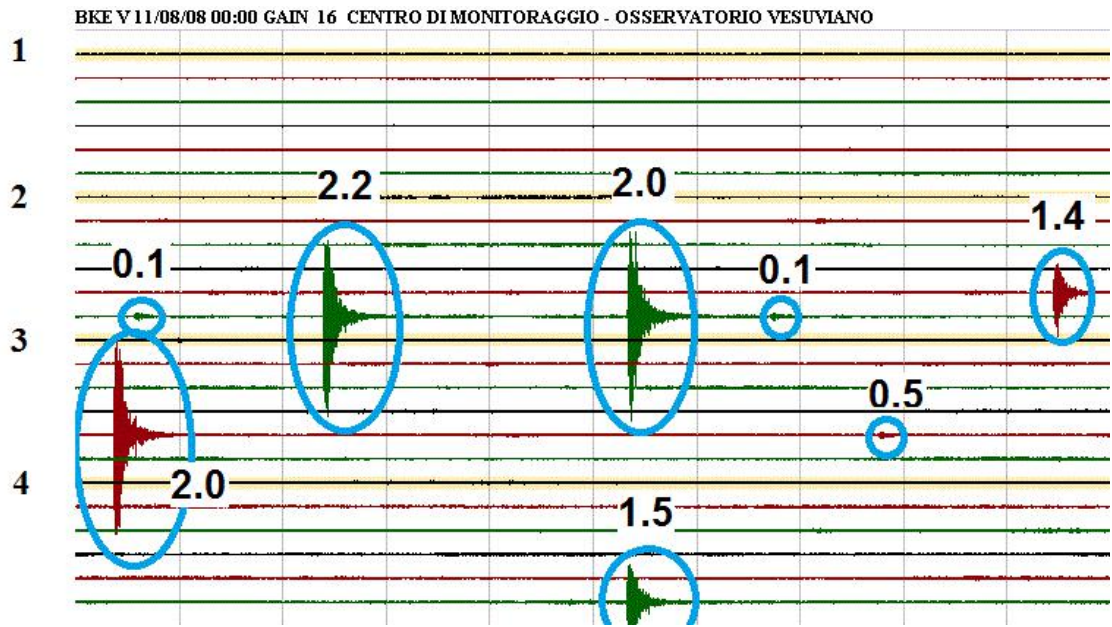

**Figure 3:** Examples of real earthquake signatures in the seismogram with relative magnitudes recorded at BKE. The data refer to the night time of Aug/08/2011 from 1:00 am to 5:00 am local time. Each line covers 10 minutes.

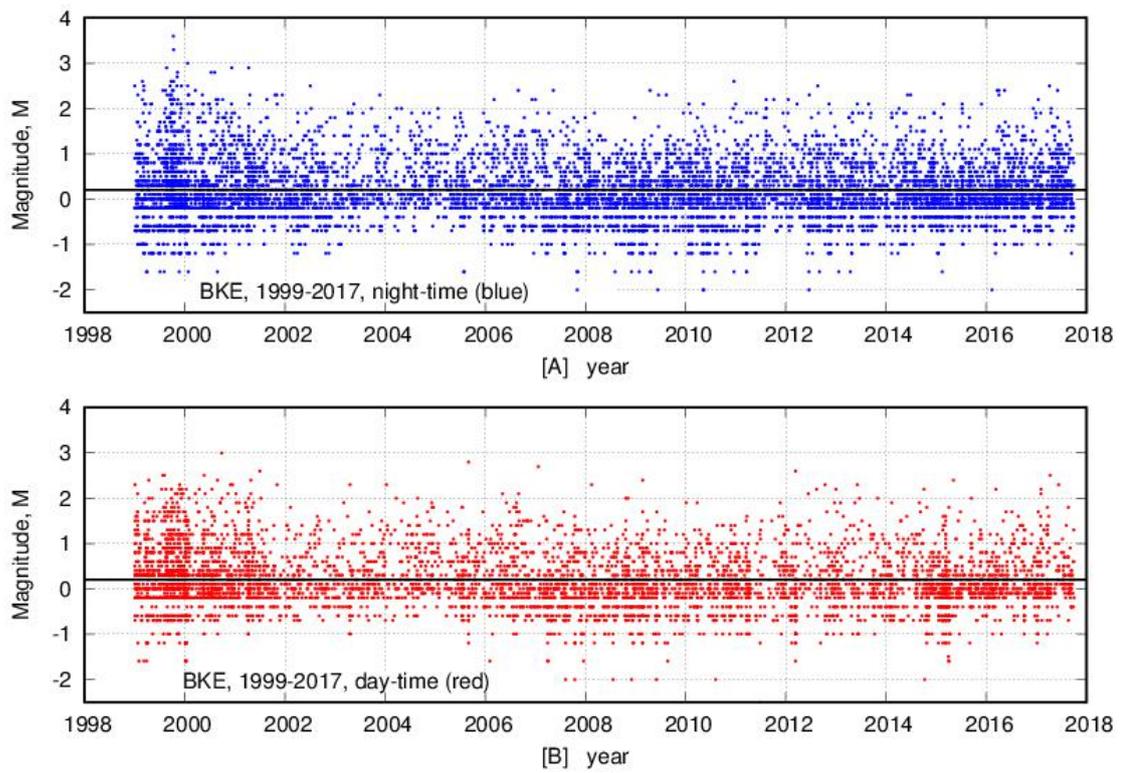

**Figure 4:** BKE catalog from from Jan/01/1999 to Oct/01/2017. The [A] blue and [B] red color refer to the night-time and day-time seismic events respectively.

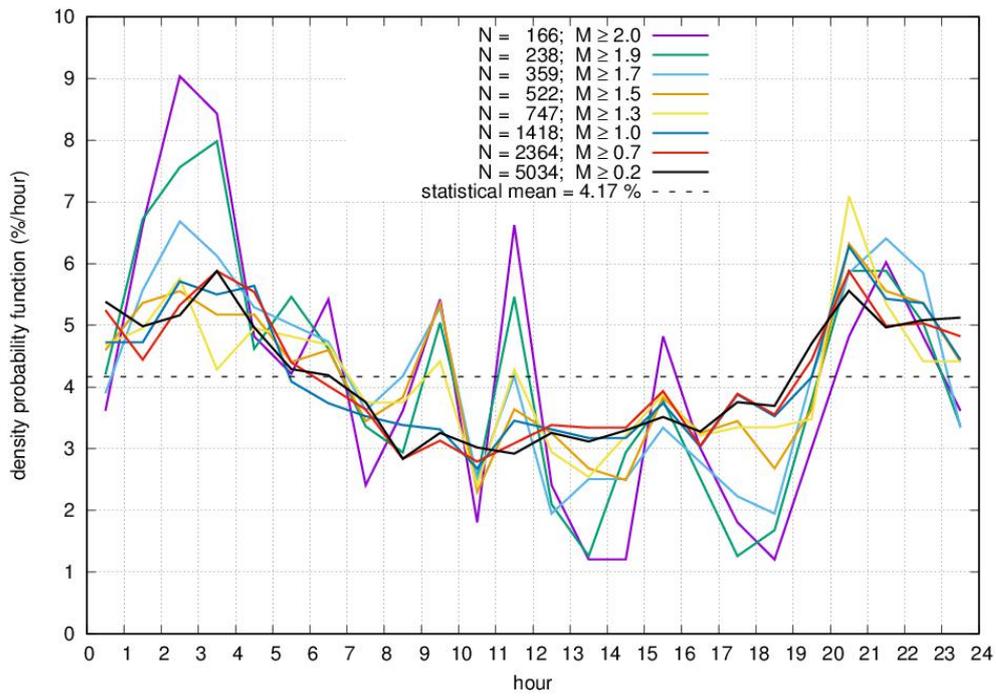

**Figure 5:** For the 1999-2014 BKE catalog. Hourly histograms of the frequency of subsets of the BKE earthquake events using different magnitude threshold levels from M ≥ 0.2 to M ≥ 2.0.

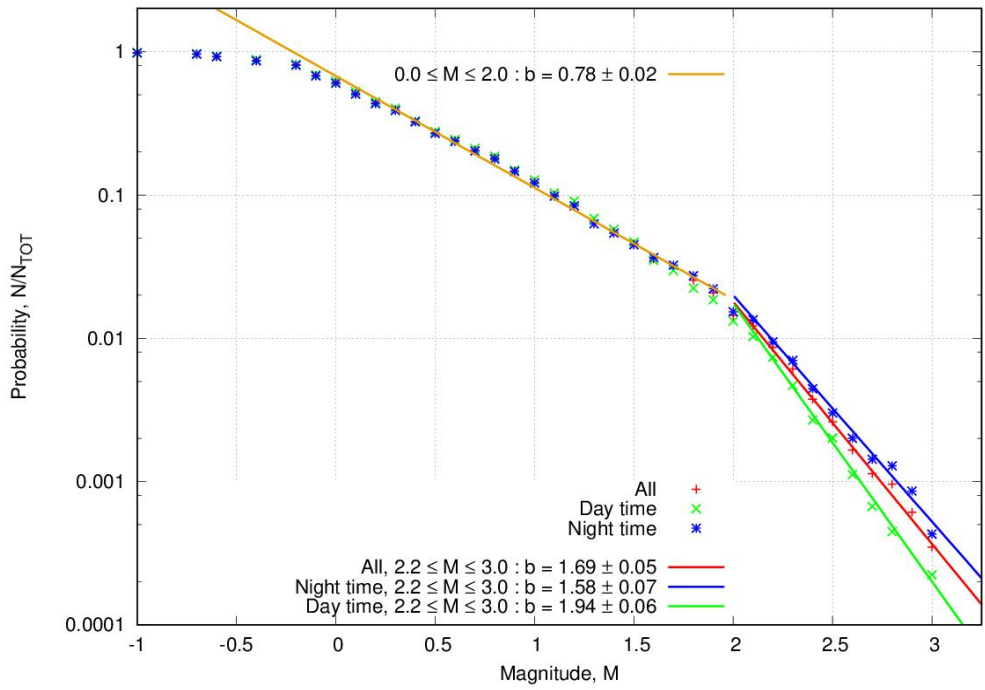

**Figure 6:** For the 1999-2014 BKE catalog. Gutenberg–Richter diagrams for all BKE events (red), for the night-time alone (blue) and for the day-time alone (green).

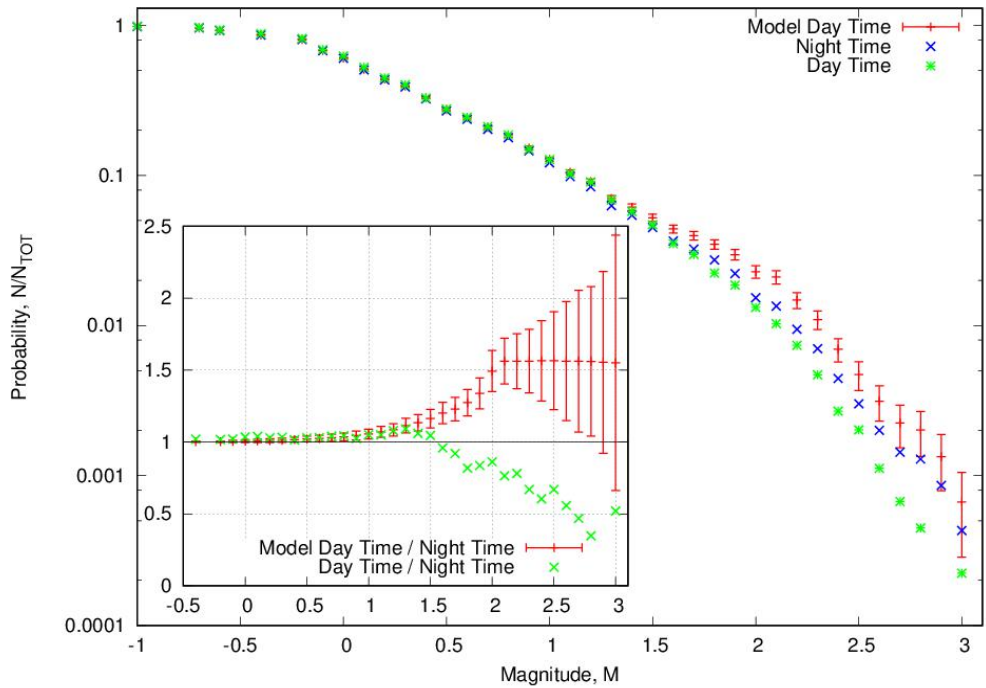

**Figure 7:** Comparison between the Gutenberg–Richter diagrams for the night-time 2events (blue), for the day-time events (green) and for the Montecarlo generated synthetic records simulating the same physical properties of the night-time catalog altered by a cultural noise effecting the events with M ≤ 2.0 (red). The insert reports the ratio between the real and synthetic day-time distribution versus the night-time distribution.

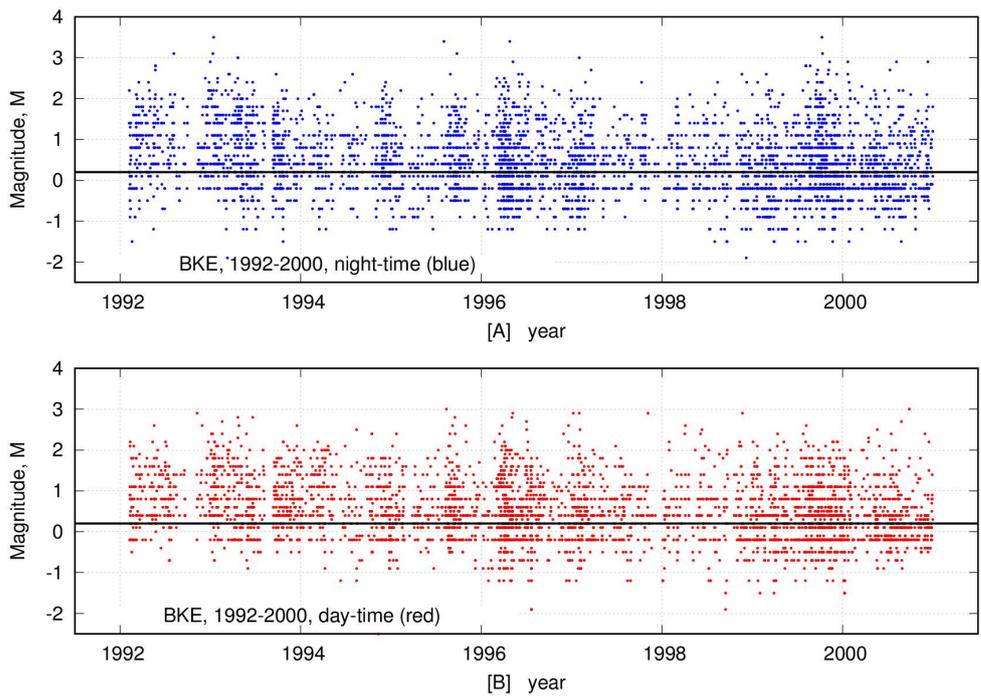

**Figure 8:** BKE catalog from from Feb/02/1992 to Dec/31/2000 (Sarao et al., 2001). The [A] blue and [B] red color refer to the night-time and day-time seismic events respectively.

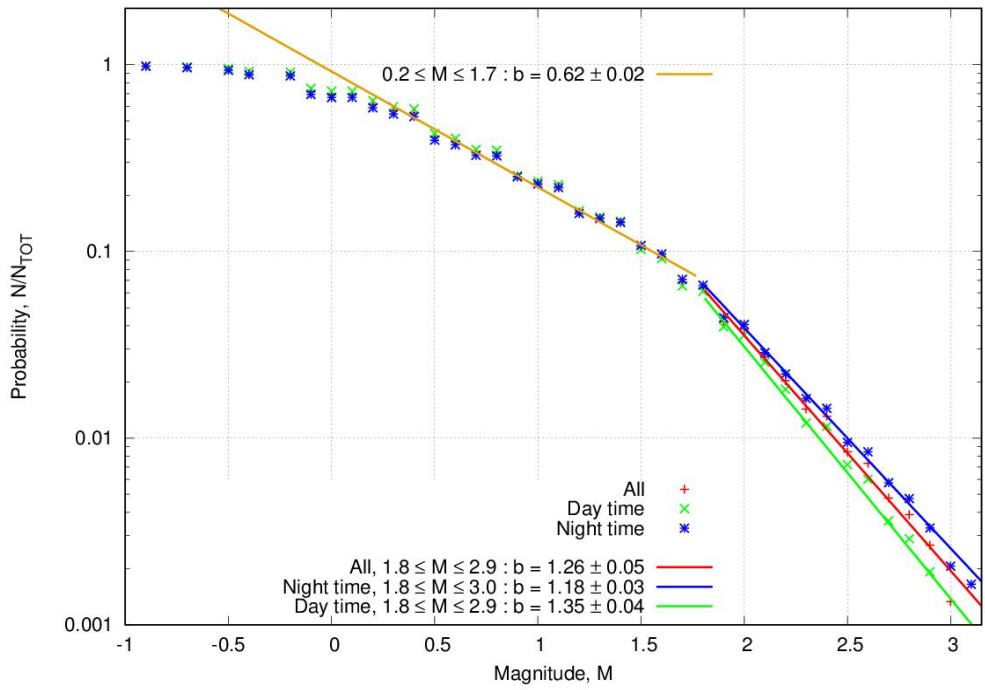

**Figure 9:** For the 1992-2000 BKE catalog. Gutenberg–Richter diagrams for all BKE events (red), for the night-time alone (blue) and for the day-time alone (green).

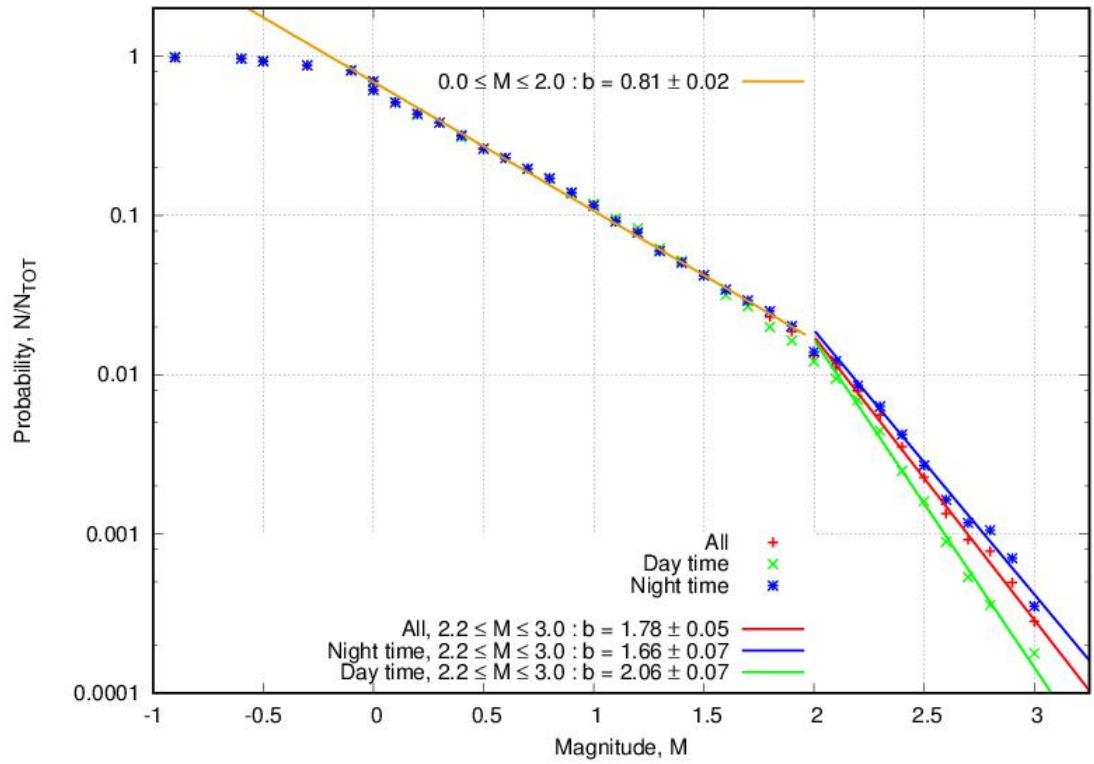

Figure 10. For the 1999-2017 BKE catalog. Gutenberg–Richter diagrams for all BKE events (red), for the night-time alone (blue) and for the day-time alone (green).